\documentclass[runningheads]{llncs}
\usepackage[T1]{fontenc}
\usepackage{graphicx}
\usepackage{booktabs}
\usepackage{amsmath}
\usepackage{subcaption}
\begin{document}
\title{Agent-Based Modelling Meets Generative AI in Social Network Simulations}
%
%\titlerunning{Abbreviated paper title}
% If the paper title is too long for the running head, you can set
% an abbreviated paper title here
%
\author{Antonino Ferraro\inst{1}\orcidID{0000-0002-1326-0325} \and
Antonio Galli\inst{1}\orcidID{0000-0001-9911-1517} \and Valerio La Gatta\inst{1}\orcidID{0000-0002-5941-4684} \and Marco Postiglione\inst{1}\orcidID{0000-0003-1470-8053} \and Gian Marco Orlando\inst{1}\orcidID{0009-0004-7136-1804} \and Diego Russo\inst{1}\orcidID{0009-0007-1095-5168} \and Giuseppe Riccio\inst{1}\orcidID{0009-0002-8613-1126} \and Antonio Romano\inst{1}\orcidID{0009-0000-5377-5051} \and Vincenzo Moscato\inst{1}\orcidID{0000-0002-0754-7696}}
\authorrunning{A. Ferraro et al.}
% First names are abbreviated in the running head.
% If there are more than two authors, 'et al.' is used.
%
\institute{Department of Electrical Engineering and Information Technology\\ University of Naples Federico II, Naples, Italy\\
\email{\{antonino.ferraro, antonio.galli, valerio.lagatta, marco.postiglione, vincenzo.moscato\}@unina.it, \\\{gian.orlando, diego.russo, giuseppe.riccio9, antonio.romano45\}@studenti.unina.it}}
\maketitle              % typeset the header of the contribution
\begin{abstract}
Agent-Based Modelling (ABM) has emerged as an essential tool for simulating social networks, encompassing diverse phenomena such as information dissemination, influence dynamics, and community formation. However, manually configuring varied agent interactions and information flow dynamics poses challenges, often resulting in oversimplified models that lack real-world generalizability. Integrating modern Large Language Models (LLMs) with ABM presents a promising avenue to address these challenges and enhance simulation fidelity, leveraging LLMs' human-like capabilities in sensing, reasoning, and behavior. In this paper, we propose a novel framework utilizing LLM-empowered agents to simulate social network users based on their interests and personality traits. The framework allows for customizable agent interactions resembling various social network platforms, including mechanisms for content resharing and personalized recommendations. 
We validate our framework using a comprehensive Twitter dataset from the 2020 US election, demonstrating that LLM-agents accurately replicate real users' behaviors, including linguistic patterns and political inclinations. These agents form homogeneous ideological clusters and retain the main themes of their community. Notably, preference-based recommendations significantly influence agent behavior, promoting increased engagement, network homophily and the formation of echo chambers. Overall, our findings underscore the potential of LLM-agents in advancing social media simulations and unraveling intricate online dynamics.

\end{abstract}

\section{Introduction} \label{sec: introduction}

Over the past decades, there has been a concerted effort among researchers and practitioners to develop computational agents capable of realistically emulating human behavior \cite{Intro_Book_Social_Science_Simulation}. Agent-Based Modelling (ABM) has emerged as a pivotal methodology for simulating intricate systems by delineating rules governing individual agents' behavior and interactions \cite{Intro_ABM}. Within the domain of social network analysis, ABM has played a crucial role in both the development and validation of novel theories pertaining to human behavior in online environments. These theories encompass a wide array of phenomena such as opinion formation \cite{ABM_Opinion_Formation}, (false) news propagation \cite{ABM_Fake_News}, and collective decision-making \cite{ABM_Collective_Decision}. Nevertheless, manually crafting agent behavior to encompass the diverse spectrum of interactions, information flow dynamics, and user engagement within social networks proves to be highly challenging. This challenge often leads to an oversimplification of agents or the social media environment itself, where underlying mechanisms are rigidly encoded in predefined parameters. Consequently, such setups are prone to researcher bias, potentially resulting in a lack of fidelity in modeling complex human behaviors, especially those involving collective decision-making \cite{ABM_Limitations}.

Modern Large Language Models (LLMs) not only excel in generating human-like text but also demonstrate remarkable performance in complex tasks requiring reasoning, planning, and communication \cite{ABM_Reasoning}. This proficiency has sparked interest in integrating LLMs with ABM, termed Generative Agent-Based Modelling (GABM). Unlike traditional ABM methods that often necessitate intricate parameter configurations, GABM leverages LLMs' capacity for role-playing, ensuring diverse agent behaviors that closely mirror real-world diversity. For instance, Park et al. \cite{Interactive_Simulacra} demonstrated that generative agents, designed for daily activities, exhibited credible individual and social behaviors, including expressing opinions and forming friendships, without explicit instructions. Similarly, Williams et al. \cite{GenAgents_Epidemic_Modelling} showcased the collective intelligence of generative agents in epidemic modeling, accurately simulating real-world behaviors like quarantine and self-isolation in response to escalating disease cases. These pioneering findings support investigating GABM as an effective approach to enhance social media simulations. To our knowledge, the seminal work by Gao et al. \cite{GenAgents_S3} lays the foundation for this research direction by qualitatively demonstrating that LLM-agents exhibit realistic behaviors related to information propagation and the manifestation of attitudes and sentiment. However, it remains unclear whether LLM-agents can accurately represent real users in terms of their personality traits (e.g., being outspoken, being critical) and interests (e.g., social issues, political preferences), regardless of the explicit emotions conveyed through their textual posts. Furthermore, their ability to exhibit community-level phenomena (e.g., homophily, polarization), as well as their susceptibility to recommendation strategies, remains uncertain.

\paragraph*{\textbf{Contributions of this work}}
In this paper, we directly target these challenges and propose a novel framework which employs LLM-empowered agents to simulate users within a social network. Initially, we construct an environment using authentic real-world social network data. To ensure the authenticity of this environment, we propose an \emph{Agent Characterization Module} that combines prompt engineering and prompt tuning to infer users' personality traits and interests. Subsequently, the simulation unfolds in two cyclical components: the \emph{Reasoning Module} that delineates each agent's decision in the simulation (e.g., posting original content, resharing, remaining inactive), and the  \emph{Interaction Module} that stores agent's past behavior and specify how agents are exposed to content from other agents (e.g., through preference-based, popularity-based or random recommendations). Notably, the Reasoning Module is fueled by the Interaction Module, offering insights for agents' informed decision-making within the simulated social network environment. We evaluate the efficacy of our proposed framework in approximating a real social media platform by scrutinizing the individual characteristics of LLM-agents compared to real users and exploring the typical network-level interactions observed in social media networks. To operationalize this objective, we formulate the following research questions (RQs):
\begin{itemize}
     \item[\textbf{RQ1:}] \emph{Do LLM-agents represent the interests of the users they are instructed to impersonate?}
    \item[\textbf{RQ2:}] \emph{Do LLM-agents form communities and/or echo chambers?}
\end{itemize}
Leveraging a large-scale Twitter dataset from the 2020 US election, we found that LLM-agents accurately mirror real users' linguistic patterns and preserve their political leaning, thus enhancing simulation authenticity. These agents also exhibit realistic behavior by resharing content aligned with their beliefs and engaging in similar communication styles as their community, reflecting online interactions accurately. Furthermore, we found that LLM-agents aggregate into homogeneous ideological groups based on their individual preferences. %and contributing to understanding echo chambers and polarization. 
Finally, we also observed the significant impact of recommendation strategies on agent behavior, emphasizing the efficacy of preference-based recommendations for promoting higher engagement and echo chambers formation. Overall, our findings prove the promising capabilities offered by LLM-agents to enhance social media simulations.

\section{Related Works} \label{sec: related}

\subsection{Agent-Based Modelling for Social Media Simulation}

The exponential progress in computational capabilities has transformed social media simulations, becoming a pivotal tool for understanding the intricate dynamics governing these digital spaces. ABM stands as a robust methodology, orchestrating interactions among individual agents based on predefined yet realistic rules. Specifically, ABM has enabled the investigation of complex online behaviors, including information dissemination \cite{Related_ABM_News_Propagation}, influence dynamics \cite{van2013agent}, and the impact of automated bots on news propagation \cite{ABM_Bot_Disinformation}. In addition, ABM has been crucial to evaluate specific disinformation countermeasures \cite{butts2023mathematical}, such as content moderation \cite{10.1145/3625007.3627489} and fake news inoculation \cite{gausen2021can}. While the above-mentioned studies highlight ABM's utility in elucidating and modeling social media phenomena, they collectively confront several limitations. First, modeling human/agent behavior often requires detailed calibration, making ABM outcomes sensitive to parameter values and assumptions/simplifications used in the simulation. Second, ABM heavily relies on predefined rules, which inherently introduce the potential for researcher bias and may impede the accurate representation of social media complexities such as the spread of multiple information narratives and/or conflicting viewpoints \cite{10148893}. Although methods like learning rules through reinforcement learning offer partial mitigation, challenges persist, particularly in scenarios where explicit reward functions for optimization are absent. In this paper, we propose a novel paradigm for social media simulations that leverages generative agents, i.e., agents empowered with LLMs' capabilities, to autonomously learn and adapt their behavior based on extensive language understanding and context reasoning, reducing the reliance on explicit parametrization and predefined rules. This paradigm shift not only enhances the fidelity and realism of the agents, but can influence the robustness and validity of ABM results.

\subsection{Generative Agents}

Modern LLMs excel not only in generating human-like text but also in complex tasks like reasoning and planning, making them valuable for enhancing simulation fidelity and complexity. The integration of LLMs with ABM, i.e., Generative Agent-Based Modelling (GABM) has garnered research interest for its potential in simulating realistic behaviors \cite{Intro_ABM_LLM_1}. For example, Park et al. \cite{Interactive_Simulacra} demonstrated that generative agents in daily activities exhibited credible behaviors at both individual and social levels without explicit instructions. Similarly, Williams et al. \cite{GenAgents_Epidemic_Modelling} showcased the collective intelligence of generative agents in epidemic modeling, mimicking real-world responses to disease outbreaks.

In this study, we contribute to the GABM research field by assessing the effectiveness of LLM-empowered agents in simulating social networks. To our knowledge, the $S^3$ framework \cite{GenAgents_S3} is the only prior work that delves into the potential of GABM for social media simulations. However, our approach differs significantly from $S^3$ in several critical aspects. First, our agent initialization strategy allows for characterizing users based on their personality and (political) interests, offering a higher level of personalization compared to $S^3$, which primarily focuses on demographic attributes (e.g., gender, age, occupation). 
Second, our Interaction Module permits custom information exposure definitions, facilitating the evaluation of diverse recommendation strategies' impacts. In contrast, $S^3$ employs a simpler and less realistic interaction mechanism where every user is uniformly exposed to all others. Third, we utilize an open-source LLM for our experiments instead of the commercial GPT3.5 service, enhancing result transparency and accessibility. Lastly, our evaluation extends the validation of LLM-agents beyond individual properties to investigate the networks they tend to form. Specifically, we move beyond news dissemination to analyze network features such as homophily, polarization, and the controversies arising from LLM-agent interactions.
\begin{figure*}[t]
    \centerline{\includegraphics[width=\linewidth]{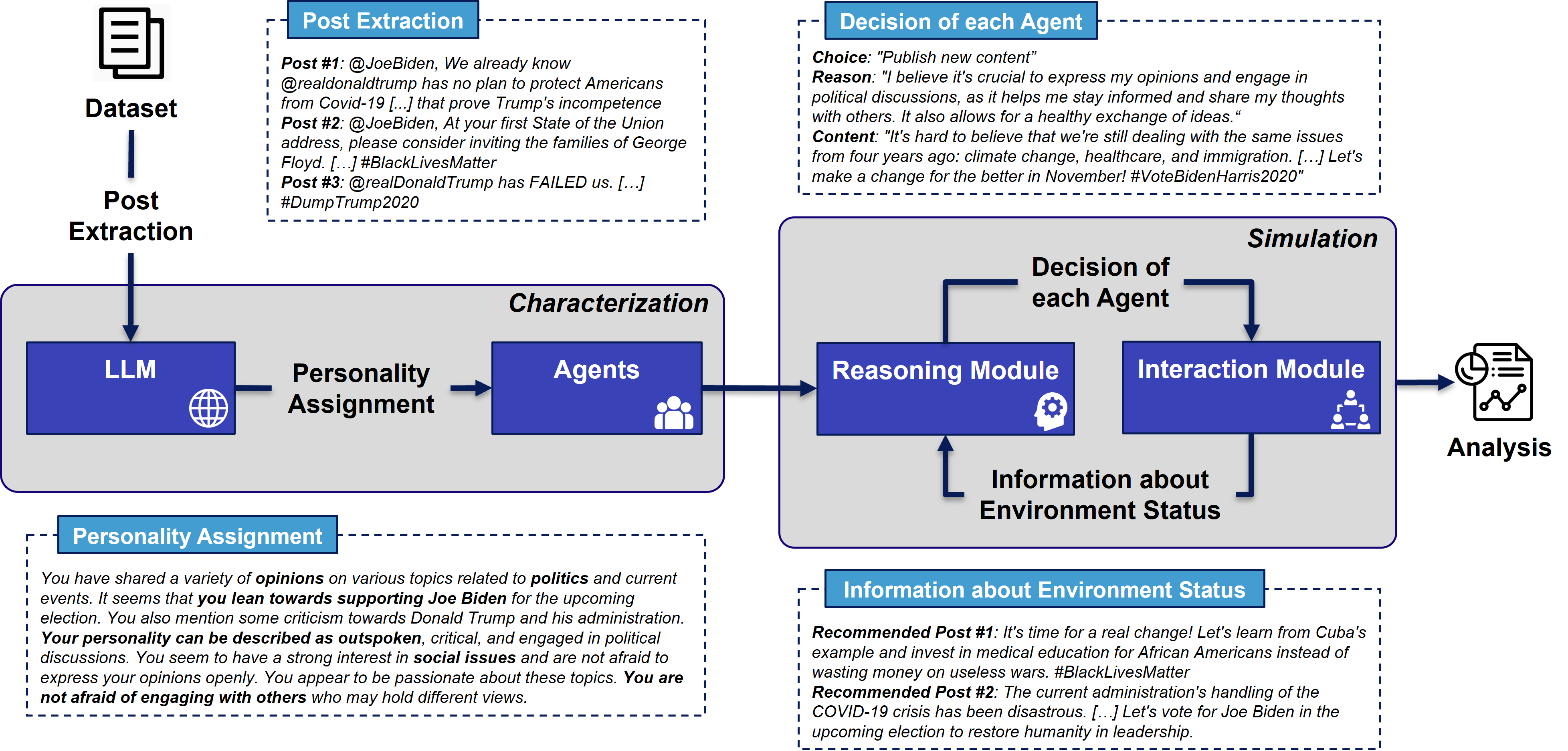}}
    \caption{Our framework comprises two primary phases: (i) \emph{Characterization}, where each agent embodies the personality traits and interests extracted (via LLM) from the original posts of the real user it is tasked to emulate; and (ii) \emph{Simulation}, where the decision-making process of each agent, represented as a \emph{Choice-Reason-Content} triple (\emph{Reasoning Module}), is stored within the \emph{Interaction Module}. Consequently, each agent autonomously makes decisions, considering the context and having access to recommended contents posted by other agents.}
    \label{Framework}
\end{figure*}

\section{Methodology} \label{sec: methodology}

Figure \ref{Framework} depicts the architecture of the proposed framework, comprising two complementary phases: (i) \emph{Characterization}, responsible for initializing generative agents based on real users' interests and personality traits, and (ii) \emph{Simulation}, responsible for executing the simulation dynamics, encompassing agents' decision-making processes and their interactions with each other. The following sections provide a detailed description of each component.

\subsection{Characterization Phase}

The primary objective of this phase is to profile each agent before commencing the simulation, where an agent simulates a user within a social network. In this context, we consider the user's original content that the agent will emulate and adopt a prompt-based approach to extract the user's personality and interests,eliminating the necessity to define apriori the parameters for agent characterization. This approach ensures diversity among individual agents, aiming to approximate the genuine interests of the respective user. Unlike prior research \cite{GenAgents_S3} focusing solely on user demographics, we conjecture that considering the user's personality and interests provides a more accurate representation of social media users' characteristics. Indeed, users personality reflects their engagement style, while their interests reveal the topics that genuinely pique their curiosity. For example, as depicted in Figure \ref{Framework} (left), an agent described as ``outspoken" and ``critical" embodies a user unafraid to voice opinions and evaluate information before engaging with it. Furthermore, its (political) interests indicate support for Joe Biden for the 2020 US election and alignment with his political stances on social issues.

\subsection{Simulation Phase}

This phase is responsible for conducting the actual simulation dynamics, including the decision of the agents and their interactions. Specifically, it unfolds in a cycle involving two modules: the Reasoning Module and the Interaction Module.

\paragraph{\textbf{Reasoning Module}}

In each iteration ($i$) of the simulation, every generative agent is immersed in an environment resembling a social media platform. In line with prior research \cite{GenAgents_S3}, we design a prompt that introduces the social media environment, highlighting the presence of other agents and outlining the possible actions to take within the environment: (i) \emph{generating original content}, (ii) \emph{resharing content from other agents}, and (iii) \emph{remaining inactive}. This setup, though simple, accurately mirrors common actions observed across various social media platforms, irrespective of platform-specific regulations. Moreover, we emphasize that these types of actions are the sole predefined aspects in the simulation. Instead, the agent behavior is not determined by deterministic or probabilistic processes as LLM-agents autonomously decide the actions and the content to generate based on their personality, interests, and other agents' behavior. Subsequently, the output of the Reasoning Module --- the response generated by each agent --- is structured as a triplet denoted by \emph{Choice-Reason-Content}. For instance, as illustrated in Figure \ref{Framework}, an agent may \emph{choose} to publish original \emph{content} to discuss persistent societal issues like climate change or immigration often targeted during electoral campaigns. Additionally, the agent provides the \emph{reasoning} behind the post, indicating an intent to engage in political discourse and foster a healthy exchange of ideas.

\paragraph{\textbf{Interaction Module}}
\begin{figure}[t]
    \centerline{\includegraphics[width=.5\linewidth]{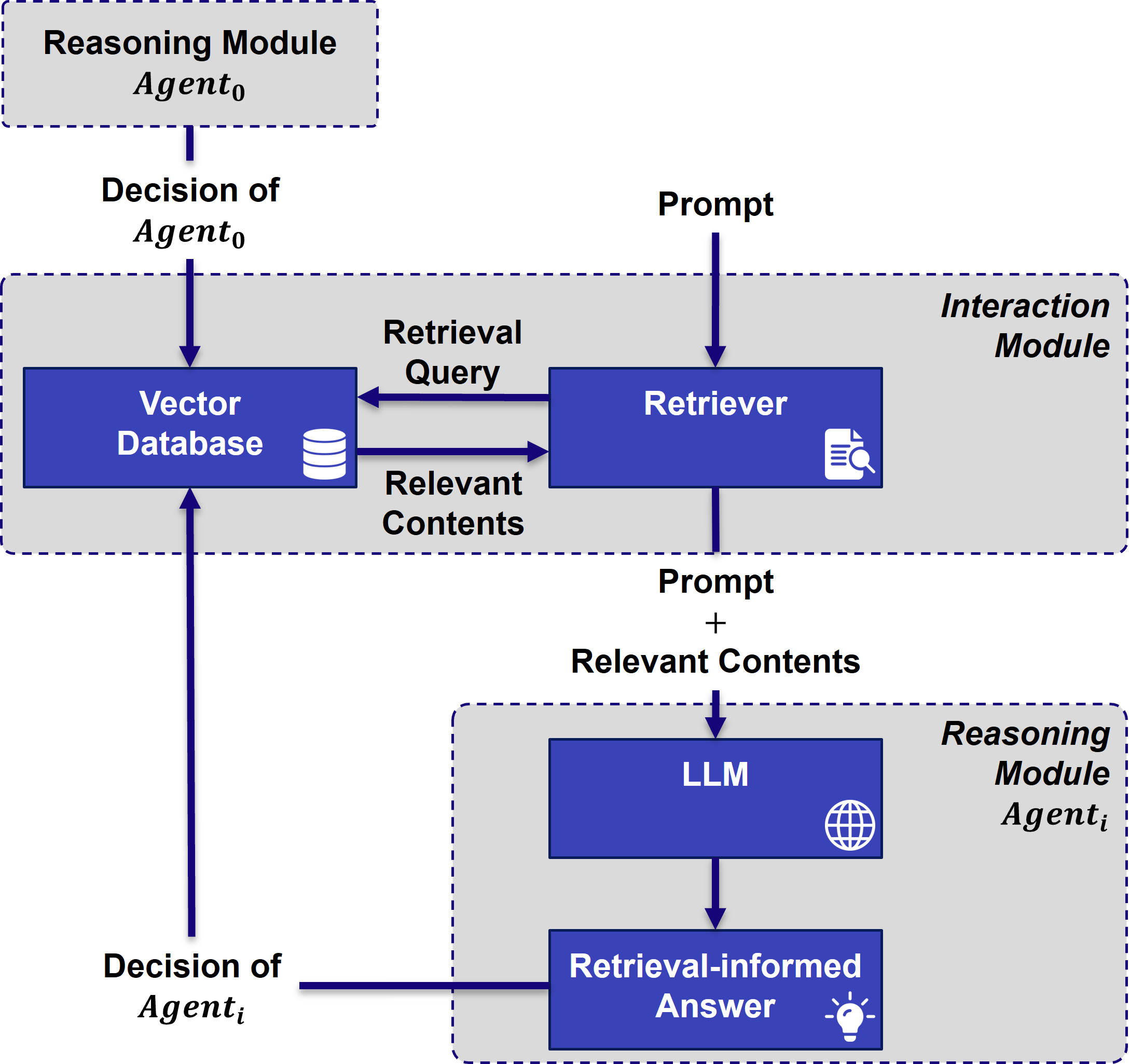}}
    \caption{
    Workflow of the RAG-empowered interaction: At the beginning of the Simulation phase, the first agent's decision is stored in the vector database. The RAG mechanism of the Interaction Module retrieves contextually relevant data from the database to enrich the prompt for the next agent to assist in making informed decisions within the Reasoning Module, taking into account the current environmental state and previously published content by other agents. The retrieval-informed decision is, in turn, stored in the vector database, repeating the cycle until the simulation ends.
    }
    \label{RAG}
\end{figure}

To achieve a comprehensive simulation of a social network, enabling interaction among agents is imperative. This involves gathering all actions performed by agents and presenting to each agent the activities of others. However, including all information generated by other users within a single prompt presents practical challenges. First, the context window of any LLM-agent is limited, causing memory saturation if all information is included in one prompt. Second, longer prompts increase the risk of hallucinations \cite{LLM_Hallucinations}, resulting in agents producing inaccurate, contextually inconsistent, or nonsensical content. To tackle these challenges, we introduce the innovative use of Retrieval-Augmented Generation (RAG). This technique enables LLM-agents to access additional contextually relevant data stored in an external vector database \cite{RAG_Technique}. Figure \ref{RAG} illustrates the integration of the RAG technique within the \emph{Interaction Module}. Specifically, the vector database is continuously updated to record all agent actions and their corresponding published contents. This tracking mechanism facilitates monitoring the simulation's progression for subsequent analysis. Furthermore, the \emph{Retriever} step within RAG serves as a recommendation system, determining which content to present to each LLM-agent. For instance, as depicted in Figure \ref{Framework}, the agent may receive recommendations for two posts from the entire corpus of published content. These posts might discuss topics such as investing in immigrant education and addressing the challenges posed by the COVID-19 pandemic. Importantly, these recommendations are aligned with the user's previous decisions, reflecting the salience of these topics during the run-up to the 2020 US election, and potentially foster further interactions between agents. Notably, this approach maintains a nearly constant prompt size throughout the simulation and enables the integration of various recommendation strategies. In our work, we focus on preference-based recommendation, i.e., recommending content aligned with agents' preferences, and random recommendation, i.e., exposing agents to random content and thus more diverse viewpoints. We will empirically assess the impacts of these strategies in our experiments.

It is important to clarify that these recommendations are not actual recommendation methods like collaborative filtering but rather a simplified approximation of them. We will empirically assess the impacts of these strategies in our experiments. Finally, the Interaction Module provides the Reasoning Module with a series of contents published by other agents to inform the LLM about the environment's state and other agents' publications, ensuring continuous updating of the environment's state regarding agents' decisions. Upon querying all agents, the current iteration concludes, and the next begins, following a round-robin logic involving sequential querying of each agent. This iterative process continues until the simulation's end, emulating social network evolution based on individual agent actions.

\section{Experiments} \label{sec: experiments}

\subsection{Experimental setup}

All experiments have been performed on a computing system featuring an 11th Gen Intel Core i7-11800H processor operating at 2.30 GHz, 16 GB of RAM clocked at 3200 MHz, and a NVIDIA GeForce RTX 3060 Laptop GPU. Each simulation comprises 10 iterations, resulting in approximately 15 hours per simulation run.  The framework\footnote{The code will be made available upon acceptance.} is implemented using PyAutogen \cite{AutoGen}, with ChromaDB\footnote{https://www.trychroma.com/} serving as the vectorial database supporting the RAG technique.  Unlike prior studies \cite{Interactive_Simulacra,GenAgents_Epidemic_Modelling} focusing on generative agents, we have chosen to employ an open-source LLM, i.e., Dolphin 2.1 Mistral 7B\footnote{https://huggingface.co/TheBloke/dolphin-2.1-mistral-7B-GGUF}, instead of proprietary foundation models (e.g., GPT-4 \cite{GPT4}, Google Gemini\footnote{https://gemini.google.com/}). This decision stems from the model's unrestricted nature and its data filtering policies designed to mitigate alignments and biases.

\subsection{Dataset}

We utilized a dataset of election-related tweets obtained through Twitter's streaming API service during the lead-up to the 2020 US election \cite{US_Dataset}. Specifically, our focus spanned six months, from June 2020 to December 2020, encompassing the latter stages of the electoral campaign and the aftermath of the election. Over this observation period, we have collected a dataset comprising more than 12 million tweets, encompassing original tweets, replies, retweets, and quotes, disseminated by 1.1 million unique users \cite{Interconnected_Nature_of_Online_Harm}. To ensure authenticity, we excluded all accounts flagged as bots by the Botometer API\footnote{https://botometer.osome.iu.edu}. Additionally, our analysis concentrated on original tweets to extract insights into users' personalities and interests. Eventually, we annotated the political affiliations of 100 users, revealing that 73 were associated with the Republican community, while the remaining 27 were aligned with the Democratic community. We have adopted this group of users to instantiate the agents of every simulation.

\subsection{Characterizing LLM-agents vs. Real Users' Interests (RQ1)}

To answer RQ1, we characterize LLM-agents and the users that they are supposed to impersonate across three dimensions:
\begin{itemize}
    \item \emph{Keywords usage}: We analyze the most relevant keywords used by LLM-agents in comparison to real users;
    \item \emph{Interests}: We examine the political leaning exhibited by LLM-agents during the simulation;
    \item \emph{Content similarity}: We investigate the semantic similarity of LLM-agents with respect to the community they belong to.
\end{itemize}

\paragraph{\textbf{Keywords usage}}

We employ YAKE \cite{Yake} and KeyBERT \cite{KeyBERT} algorithms to extract keywords using statistical features and contextual embeddings, respectively. Specifically, we focus on original content and merge the vocabulary used by all agents during simulation as well as the vocabulary used by real users. This approach is motivated by the significant diversity in the scale of simulation activities compared to real users, where real users publish much more original tweets than their LLM-agents' counterparts. Table \ref{tab: Main Keywords} shows the top-10 keywords used by LLM-agents and real Twitter users, demonstrating that the simulation, and consequently the LLM-agents, align with the primary discussion topics of real Twitter conversations, particularly the debates between Trump and Biden during the 2020 US election. Additionally, Figure \ref{Distribution of Keywords} shows the distributions of keywords usage in both real Twitter discussions (in blue) and the simulation (in red). The visibly right-skewed distributions, coupled with statistical validation through a Mann–Whitney test ($p$-value$<0.05$), affirm the simulation's fidelity in reflecting natural language patterns. This observation underscores the simulation's ability to accurately mimic the overarching themes of Twitter conversations, with few keywords being frequently utilized while others are less prevalent. Collectively, these results demonstrate that LLM-agents effectively capture the main topic of the conversation, indicating the robustness of the simulation in replicating real-world discourse.
\begin{table}[t]
    \centering
    \begin{minipage}{.3\textwidth}
        \centering
        \caption{Top-10 Keywords in Real Case and Simulation}
        \label{tab: Main Keywords}
        \begin{tabular}{cc} \toprule 
            \textbf{Real Case} & \textbf{Simulation} \\ \midrule
            realdonaldtrump & trump \\
            trump & president \\
            president & biden \\
            biden & administration \\
            joebiden & freedom \\
            people & maga\\
            america & actions\\
            covid & change\\
            time & covid\\
            maga & state\\ \bottomrule
        \end{tabular}
    \end{minipage}%
    \hfill
    \begin{minipage}{.5\textwidth}
        \centering
        \includegraphics[width=\linewidth]{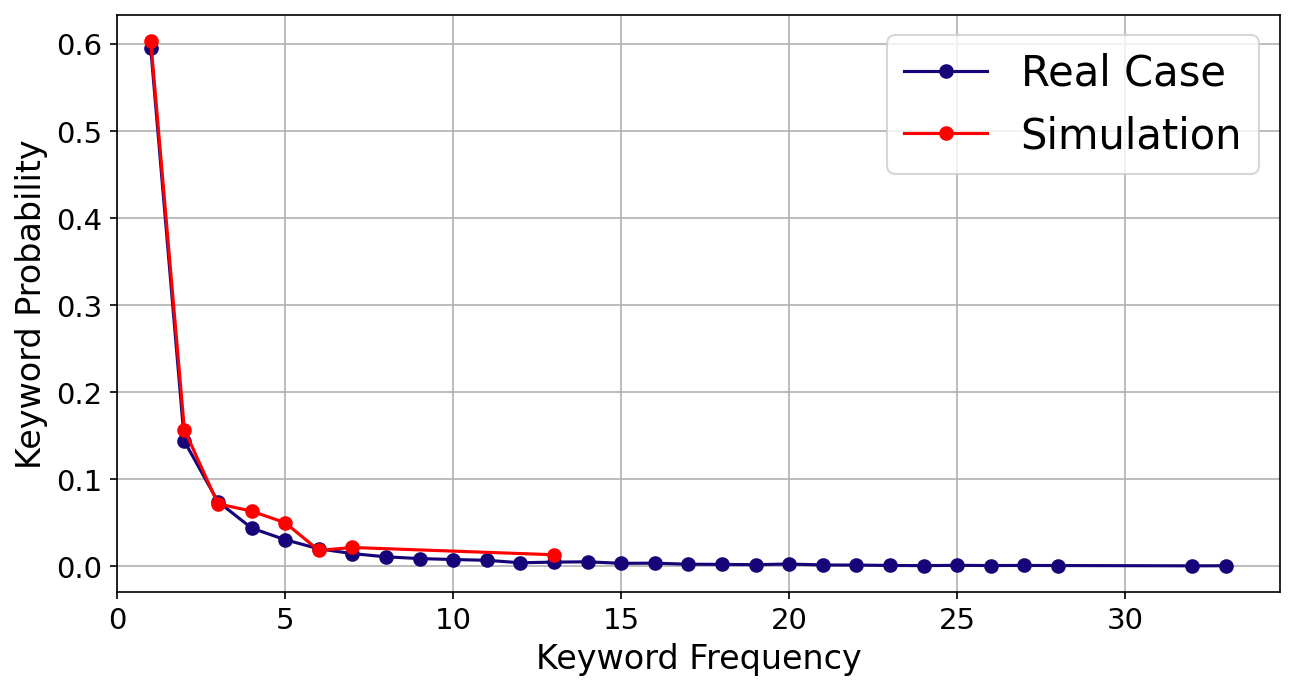}
        \captionof{figure}{The distributions of keywords usage in real Twitter discussions and the simulation. The x-axis represents the number of occurrences of a keyword, while the y-axis represents the probability of encountering a keyword with a specific frequency.}
        \label{Distribution of Keywords}
    \end{minipage}
\end{table}

\paragraph{\textbf{Interests}}

We examine the political leaning of LLM-agents as a proxy for their (political) interests. Initially, we fine-tuned a BERT-base transformer using the annotated dataset of 100 Twitter users, achieving impressive performance metrics: 91\% accuracy and 94\% F1 score. Subsequently, we utilized this model to predict the political leaning scores of LLM-agents at the simulation's conclusion. As previously mentioned, we investigate two exposure strategy: preference-based recommendation and random recommendation. To enhance robustness, we repeated the random recommendation simulation three times. Notably, the RAG-enhanced Interaction Module allows a seamless integration of any recommendation strategy. Table \ref{tab:Results of Political Alignment Analysis and Interaction Patterns} shows a strong positive correlation between the political leaning scores of real users and LLM-agents\footnote{We employed the Spearman index with a significance level of $p$-value$<0.05$ to verify this correlation.}. Specifically, the majority of agents (always $\geq86\%$) has retained the political orientations of the real users they are instructed to impersonate. Interestingly, this is irrespective of the type of recommendation strategy. However, fewer agents alter their alignments during the simulations. After qualitative scrutiny, we have found that these shifting agents typically impersonate nonpartisan users. Therefore, exposure to diverse ideas may blur their political orientations. Altogether, this analysis indicates that LLM-agents accurately interpret real users' political alignments, regardless of the content they are exposed to.
\begin{table}[t]
    \small
    \centering
    \caption{Results of Political Leaning Analysis and Interaction Patterns}
    \label{tab:Results of Political Alignment Analysis and Interaction Patterns}
    \setlength{\tabcolsep}{3pt} % Increase space between columns
    \resizebox{\textwidth}{!}{%
    \begin{tabular}{lcccc|ccc} \toprule
        & \multicolumn{4}{c|}{\textbf{Political Leaning Analysis}} 
        & \multicolumn{3}{c}{\textbf{Interaction Patterns}} \\ \cmidrule{2-8}
        & \begin{tabular}[c]{@{}c@{}}\textbf{Consistent}\\ \textbf{Users}\end{tabular} 
        & \begin{tabular}[c]{@{}c@{}}\textbf{Changed}\\ \textbf{Users}\end{tabular} 
        & \begin{tabular}[c]{@{}c@{}}\textbf{Spearman}\\ \textbf{Coefficient}\end{tabular} 
        & \textbf{P-Value}
        & \begin{tabular}[c]{@{}c@{}}\textbf{Original}\\ \textbf{Publications}\end{tabular} 
        & \begin{tabular}[c]{@{}c@{}}\textbf{Non}\\ \textbf{Interactions}\end{tabular} 
        & \begin{tabular}[c]{@{}c@{}}\textbf{Reshares}\end{tabular} \\ \midrule
        Pref-Based & 88\% & 12\% & 0.494 & <0.05 & 38.9\% & 10.5\% & 50.6\% \\
        Random 1 & 86\% & 14\% & 0.498 & <0.05 & 90.0\% & 1.6\% & 8.4\% \\
        Random 2 & 88\% & 12\% & 0.490 & <0.05 & 89.1\% & 2.5\% & 8.4\% \\
        Random 3 & 89\% & 11\% & 0.509 & <0.05 & 90.1\% & 1.9\% & 8.0\% \\ \bottomrule
    \end{tabular}
    }
\end{table}

\paragraph{\textbf{Content Similarity}}

\begin{figure}[t]
    \centering
    \begin{minipage}{0.45\textwidth}
        \centering
        \includegraphics[width=\linewidth]{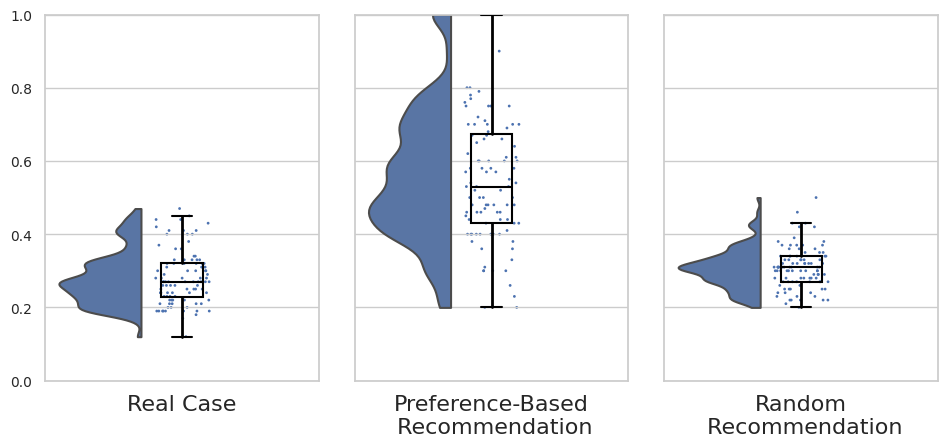}
        \caption{Comparison of self-similarity distributions between real users and LLM-agents.% under preference-based and random recommendation settings.
        }
        \label{Raincloud Plot for Self-Similarity}
    \end{minipage}
    \hfill
    \begin{minipage}{0.45\textwidth}
        \centering
        \includegraphics[width=\linewidth]{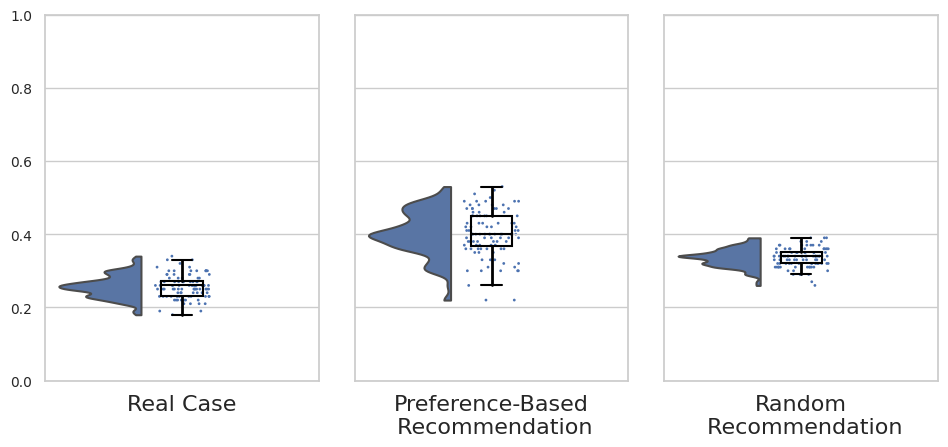}
        \caption{Comparison of intra-cluster similarity distributions between real users and LLM-agents. % under preference-based and random recommendation settings.
        }
        \label{Raincloud Plot for Intra-Cluster Similarity}
    \end{minipage}
\end{figure}

Lastly, we investigate the semantic similarity of the content published by LLM-agents. Specifically, we utilized a sentence-transformers model\footnote{https://huggingface.co/sentence-transformers/all-MiniLM-L6-v2} to extract contextual embeddings for each original content posted by LLM-agents. We then constructed a cosine similarity matrix $\mathcal{M} = \{m_{ij} = \text{sim}(c_i, c_j)\}$, where $c_i$ and $c_j$ represent the vector embeddings of two posts and $\mathcal{M} \in \mathcal{R}^{k \times k}$, $k$ being the total number of posts published by the agents. To evaluate the semantic similarity, we define two measures:
\begin{itemize}
    \item \emph{Self-Similarity}: it assesses the similarity between posts from the same agent $A$. In other words, it represents the average cosine similarity measured on $\mathcal{M}$ considering only the agent's original content, that is $\{m_{ij} \} | c_i \ne c_j \wedge c_i, c_j$ posted by $A$;
    \item \emph{Intra-Cluster Similarity}: it assesses the similarity between posts from an agent $A$ and those from other agents within the same political group $\mathcal{A}_p=\{ A_1, A_2, \cdots, A_k | A_i \ne A \wedge pol(A_i) = pol(A) \}$, $pol$ being the political leaning scoring function. Concretely, the metric is determined by averaging the cosine similarity of $\mathcal{M}$'s elements, that is $\{m_{ij} \} | c_i$ and $c_j$ posted by $A$ and $A_i \in \mathcal{A}_p$, respectively.
\end{itemize}
Figure \ref{Raincloud Plot for Self-Similarity} shows the distributions of \emph{Self-Similarity} for real users and LLM-agents across two simulation scenarios: one featuring preference-based recommendation and the other with random recommendation. Notably, for the real case, the cosine similarity matrix is built considering the original tweets of the users. We notice a clear disparity in self-similarity  between LLM-agents and real users within the preference-based recommendation framework. Specifically, the median self-similarity for real users (0.274) contrasts with that of LLM-agents (0.531), with the latter exhibiting significantly higher self-similarity\footnote{This finding was validated using a Mann–Whitney test ($p$-value$<0.05$)}. This observation indicates a tendency for the same LLM-agent to converge more strongly around similar topics compared to the real user it is impersonating, suggesting that preference-based recommendation might penalise the ability of LLM-agents to engage with brand-new content. Conversely, random recommendations appear to mitigate this polarization trend and, surprisingly, achieve a better approximation of the real case. Additionally, Figure \ref{Raincloud Plot for Intra-Cluster Similarity} replicates the same analysis but considers \emph{Intra-Cluster Similarity}, revealing minimal distinctions among the three distributions. Surprisingly, the simulation employing preference-based recommendation aligns better with the real scenario compared to the previous analysis. This suggests that although the agents may not publish as diverse content as their real counterparts, they still preserve the overall semantic characteristics of their community. This outcome, combined with the previous discovery that the majority of LLM-agents retain the political leaning of their real counterparts, implies an accurate representation of the target users and an effective portrayal of their respective communities by the LLM-agents.

\subsection{Community Formation and Echo Chambers Among LLM-Agents (RQ2)}

\begin{figure}[t]
    \centerline{\includegraphics[width=.7\linewidth]{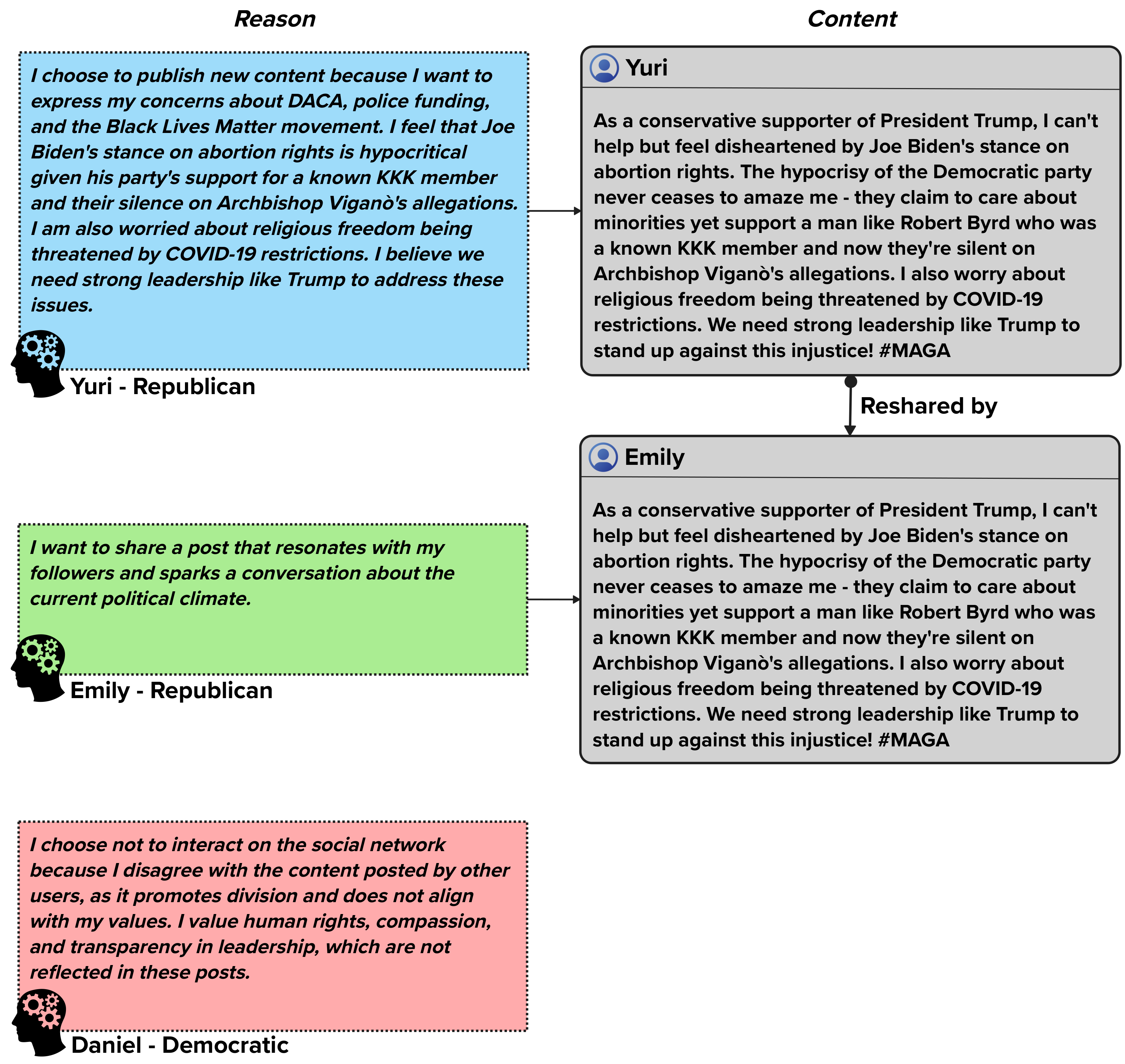}}
    \caption{Distinct behavioral reactions of agents to content based on individual preferences and political alignment can be observed. Yuri contributes by posting original content aligned with its preferences. Emily actively engages with followers by resharing content. Conversely, Daniel refrains from interaction, influenced by exposure to content conflicting with its preferences.}
    \label{Example of output}
\end{figure}

To address RQ2, we delve into agent behavior, specifically focusing on interactions, i.e., reshares, made by LLM-agents during the simulation. Figure \ref{Example of output} depicts an illustrative example involving three agents --- Yuri, Emily, and Daniel --- engaging in a political discussion regarding the 2020 US election. Daniel, a Democrat, refrains from further social media activity due to conflicting views with other agents. Conversely, Yuri posts new content highlighting policy concerns about abortion rights and COVID-19 restrictions. In an effort to engage in the ongoing conversation with her followers, Emily opts to reshare Yuri's post, given their shared political cluster and similar concerns. To quantitatively assess agents' behavior, we formalize the \emph{interaction graph} as follows: the nodes represent LLM-agents, and the (directed) edges represent resharing activity from the publisher agent to the agent that performs the reshare. Subsequently, we delve into the agents' resharing activity and examine community-based metrics, including homophily and controversy.

\paragraph{\textbf{Sharing Activity}}

Table \ref{tab:Results of Political Alignment Analysis and Interaction Patterns} shows the interaction patterns in terms of the percentages of actions performed by the agents. The influence of the recommendation system is evident: in simulations featuring random recommendations, agents exhibit a significant decrease in content resharing from others ($\leq8.4\%$) compared to simulations employing preference-based recommendations (50.6\%). This trend stems from the increased exposure to (random) content not aligned with the agent's preferences. Indeed, the random recommendation strategy results in agents publishing original content approximately 90\% of the times in an attempt to push their ideas in the environment. Surprisingly, while personalized recommendations usually boost user engagement on real social media platforms \cite{Recommender_User_Engagement}, our simulations reveal a greater proportion of non-interactions when recommendations are preference-based (10.5\%) with respect to random recommendation ($\leq2.5\%$). We attribute this to our empirical observation that the preference-based recommendation is not flawless and may sometimes suggest irrelevant posts to the agents. Overall, these findings suggest that the recommendation strategy strongly affects the agents' activity, with preference-based recommendations fostering increased interactions among agents. 

\paragraph{\textbf{Homophily}}

To investigate homophily, we examine the above-mentioned \emph{interaction graph} among agents in each simulation. Homophily, defined as individuals' inclination to associate with others sharing similar attributes, leads to the formation of homogeneous groups \cite{Homophily_Links}. Given the political context of our data, we focus on agents' political alignment to delineate group clusters. Subsequently, we analyze the inter-cluster edges of the interaction graph, representing connections between nodes from different clusters. 
The results of the homophily analysis presented in Table\ref{tab:Results of Homophily Analysis and Controversy Metrics}  indicate that networks from simulations employing preference-based recommendations exhibit reduced inter-cluster connectivity, signifying a stronger homophily effect. This results in a 10\% lower number of inter-cluster edges than the number needed to declare the network non-homophilic \cite{kim2017effect}. Statistical analyses confirm the significance of these homophily values (\textit{p-value $<$ 0.05}). Conversely, simulations using random recommendations demonstrate relatively higher inter-cluster connectivity (+4\% inter-cluster edges) with respect to preference-based recommendation, resulting in networks that are not statistically homophilic (\textit{p-value $>$ 0.05}). These results affirm our previous findings regarding the determining effect of the recommendation strategy, but also complement them by illustrating that preference-based recommendation not only fosters agents' engagement but also encourages the formation of distinct clusters with limited inter-cluster interaction.
\begin{table}[t]
    \centering
    \caption{Results of Homophily Analysis and Controversy Metrics}
    \label{tab:Results of Homophily Analysis and Controversy Metrics}
    \setlength{\tabcolsep}{3pt} % Increase space between columns
    \resizebox{0.85\textwidth}{!}{%
    \begin{tabular}{lcc|ccc} \toprule
        & \multicolumn{2}{c|}{\textbf{Homophily Analysis}} 
        & \multicolumn{3}{c}{\textbf{Controversy Metrics}} \\ \cmidrule{2-6}
        & \textbf{Modularity}
        & \begin{tabular}[c]{@{}c@{}}\textbf{Homophily?}\end{tabular} 
        & \textbf{RWC}
        & \textbf{BCC}
        & \textbf{GMCK} \\ \midrule
        Pref-Based & 0.375 & \begin{tabular}[c]{@{}c@{}}Yes, -10\% inter-cluster edges\\ (p-value <0.05)\end{tabular} & \textbf{0.692} & \textbf{0.423} & \textbf{0.334} \\
        Random 1 & 0.416 & No, +4\% inter-cluster edges & 0.423 & 0.189 & 0.071 \\
        Random 2 & 0.416 & No, +6\% inter-cluster edges & 0.541 & 0.230 & 0.268 \\
        Random 3 & 0.428 & No, +4\% inter-cluster edges & 0.431 & 0.362 & -0.009 \\ \bottomrule
    \end{tabular}
    }
\end{table}

\paragraph{\textbf{Echo Chamber Analysis}}

In line with previous works \cite{SNA_Echo_Chamber_Definition,Controversy_1,SNA_Quantifying_Controversy}, we utilize established metrics to examine the emergence of echo chambers in our simulations. These metrics analyse the \emph{interaction graph} to evaluate the level of controversy within discussions. Specifically, we utilize Random Walk Controversy (RWC) to analyze transition probabilities between ideological clusters, Betweenness Centrality Controversy (BCC) to assess partition distances, and Boundary Connectivity (GMCK) to compute the structural arrangement of the interaction graph \cite{SNA_Quantifying_Controversy}. In all cases, the higher the metric, the more controversy is the \emph{interaction graph}. The underlying premise is that contentious topics often involve individuals with contrasting viewpoints engaging in dialogue, while individuals sharing similar beliefs tend to reinforce each other's arguments \cite{Controversy_1}. Results in Table \ref{tab:Results of Homophily Analysis and Controversy Metrics} unveil a notable prevalence of controversy in preference-based simulations, indicating a stronger inclination towards echo chamber formation. Conversely, random recommendation mitigates echo chamber formation, promoting a broader diversity of opinions. These findings underscore that LLM-agents within a community are unlikely to interact with individuals holding opposing viewpoints, suggesting that the recommendation strategy not only influences community formation but also reinforces these communities by fostering closed networks, wherein users are exposed to limited diversity.

\section{Conclusions \& Future Works} \label{sec: conclusions}

In this paper, we proposed a novel framework that integrates agent-based modeling with LLM capabilities for social media simulation. Our framework incorporates a Characterization Module, enabling the inference of realistic users' personality traits and interests. Furthermore, our Interaction Module pioneers the application of RAG mechanisms to implement various recommendation strategies. By focusing on Twitter discussions surrounding the 2020 US election, we demonstrated that the simulated LLM-agents effectively mirror the users they are tasked to emulate, maintaining their original political orientations and preserving the thematic content within their respective communities (RQ1). Furthermore, our exploration into their interactions has unveiled a tendency for these agents to cluster with like-minded peers, especially under preference-based recommendation settings (RQ2). %, wherein agents reinforce their pre-existing beliefs .

Moving forward, our research aims to explore several key directions. First, we plan to augment our framework with more sophisticated recommendation algorithms, e.g., based on collaborative filtering. Additionally, we aim to enable LLM-agents to perform a broader spectrum of actions (e.g., following and liking posts). %Second, we seek to explore how LLM-agents can simulate complex social media behaviors, including cross-platform dynamics and coordinated influence campaigns. 
Second, we are planning to expand our research to include multiple LLMs, which will allow us to analyze potential biases across different models. Lastly, we emphasize the flexibility of our framework beyond social media simulations, highlighting potential cross-disciplinary applications in simulation fields (e.g., epidemiology or business planning). 

% Credits and Acknowledgments section
\begin{credits}

\subsubsection*{\ackname}
This work was supported by the European Project DEUCE, Digitalising European Uncontested Claims Enforcement. Grant Number: 101138437. Call: JUST-2023-JCOO. Type of Action: JUST-LS.

\end{credits}

%. Exploring applications in other domains, such as epidemiology or business planning, would further enhance its relevance and impact.
%
% ---- Bibliography ----
%
% BibTeX users should specify bibliography style 'splncs04'.
% References will then be sorted and formatted in the correct style.
%
\bibliographystyle{splncs04}
\bibliography{bibliography}
\end{document}